# Buckling Bars and Boxy Bulges

M. R. Merrifield

*Department of Physics, Southampton University, Southampton, SO17 1BJ, UK*

**Abstract.** It has been suggested that the peanut-shaped bulges seen in some edge-on disk galaxies are produced when bars in these galaxies buckle. This paper reviews the modelling which seeks to show how bars buckle, and I present a very simple new model which captures the essential physics of this process. I then discuss the problems in establishing observationally the connection between peanut-shaped bulges and bars: confirmation of the link has proved difficult because boxy bulges are only apparent in edge-on galaxies whereas bars are only easily detectable in more face-on systems. Finally, I present a new technique which avoids this difficulty by searching for the distinctive kinematic signature of an edge-on bar; application of this method to spectra of peanut-shaped bulges reveals that they are, indeed, associated with hidden bars.

## 1. Introduction

Around 30% of disk galaxies are observed to contain central bar-like structures (Sellwood & Wilkinson 1993). Normal mode analysis (e.g. Hunter 1992) and N-body simulations (e.g. Sparke & Sellwood 1987) show that a global instability in self-gravitating disks provides a natural explanation for the formation of such bars. As will be shown in Section 2., calculations also predict that thin bars are themselves unstable, and they should rapidly buckle perpendicular to the plane of the disk, ultimately producing a thickened bar which is peanut-shaped when viewed from the side.

Photometric work by Shaw (1987) has shown that at least 20% of edge-on disk galaxies have such peanut-shaped bulge isophotes. This percentage is sufficiently close to the fraction of galaxies with bars for it to be tempting to conclude that all the peanut-shaped bulges that we see are produced by central bars. Unfortunately, as will be seen in Section 3., the link between these phenomena has eluded direct confirmation. Fundamentally, the problem is that any peanut shape to the bulge is only detectable in a galaxy which lies close to edge-on, whereas a bar can only be seen easily in a more face-on system.

Section 4. shows how this problem can be overcome by using kinematic observations to probe the full three-dimensional structure of disk galaxies. To this end, in collaboration with Konrad Kuijken (Kapteyn Instituut) I have obtained spectra of edge-on galaxies with peanut-shaped bulges. These observations reveal the characteristic kinematic signature of bars at the centres of these systems, providing the first strong observational evidence that peanut-shaped bulges are a by-product of bar formation.

## 2. The Buckling Instability

The dramatic bar instability in cold galaxy disks has made such systems very popular with N-body simulators. Due to computational limitations, most early N-body models were restricted to two dimensions. However, ground-breaking work by Combes & Sanders (1981) indicated that important physics was being missed through this restriction: their fully three dimensional N-body simulations of a bar-unstable disk galaxy showed that the bar thickens out of the plane to produce a distinctive peanut shape. More recently, with the advent of much more powerful computers, this phenomenon has been the subject of detailed investigation (Combes et al 1990, Raha et al. 1991). This work has shown that, once a bar forms, it rapidly bends out of the plane of the disk forming initially a boomerang shape and ultimately thickening up to produce a boxy or peanut-like structure. Combes et al. (1990) and Pfenniger & Friedli (1991) have examined the three-dimensional orbits in barred potentials, and they have shown that stars on orbits lying close to a 2:1 vertical resonance are unstable, and the orbits end up bending out of the plane. They therefore proposed that this instability is responsible for the vertical buckling of bars.

Merritt & Sellwood (1994) have pointed out that this orbital instability cannot provide the full story. Firstly, the gravitational potential changes as the bar buckles, and so considering the stability of orbits in a fixed potential may not provide an accurate model of the system. Secondly, not all stars lie on the vertically unstable orbits, and yet all the stars in the bar follow the bend. Individual stellar orbits cannot, therefore, be treated in isolation — the buckling instability can only be fully understood as a collective phenomenon.

In the light of these comments, Merritt & Sellwood went on to describe a very general criterion for assessing the vertical stability of bars: a star travelling along the major axis of a bar will have a natural frequency of oscillation perpendicular to the bar, $\Omega_z$. If the bar is slightly bent, then its gravitational potential will be slightly perturbed, and a star moving back and forth along the bar with a frequency $\Omega_x$ will encounter this perturbation as a periodic forcing with frequency $\Omega_{\text{force}} = 2\Omega_x$. From the theory of forced harmonic oscillators, we know that a particular star will respond in phase with the forcing term if the forcing frequency is less than or equal to its natural frequency of oscillation, and out of phase if this inequality is not met. Stars responding in phase with the forcing term will make the bend grow to a larger amplitude, and so the overall system will be unstable if

$$2\Omega_x \leq \Omega_z \tag{1}$$

for the majority of the stars. Since a thin bar produces a strong gravitational restoring force toward the plane, $\Omega_z$ is large and so inequality (1) is likely to be met and the bar will buckle. Notice also that the largest response to a perturbation will occur when the gravitational forcing term is close to resonance with the natural frequency ($2\Omega_x \approx \Omega_z$), as is also implied by the orbital instability analysis.

Figure 1.   Edge-on view of a "shoe box" bar, showing the effects of bending such a system when it is populated by stars whose orbits lie close to the 2:1 vertical resonance. For clarity, only stars moving from left to right have been shown.

   One shortcoming in the analysis that led to this criterion is that it is only strictly valid for stars whose unperturbed paths run along the major axis of the bar. When a bar is bent, stars on non-major axis orbits will also respond in phase with the forcing gravitational perturbation if they meet inequality (1). However, some such orbits will experience a perturbation where the force is in the opposite direction to the bend, and so stars on these orbits will try to flatten the bar back out, opposing the instability. It is therefore not immediately clear what the collective behavior of all the stars will be in response to a bend in the bar.
   Ultimately, if we are to understand the cause of the bar buckling instability, we need a model which recognizes that bars are intrinsically three-dimensional structures with different stars following different orbits. We also need to be able to treat explicitly the collective response of all the stars on their particular orbits in order to investigate the instability. A particularly simple model which meets these criteria is the "shoe box" bar. In this model, stars are treated as free-streaming particles which bounce elastically from the confining "walls" of a rectangular box. We thus treat the self-gravity of the bar impulsively via the interactions with the walls, and we can investigate the tendency of the system to buckle by measuring the difference between the pressures exerted on the top and bottom of the box due to the stellar "collisions" with these surfaces. For example, Fig. 1 shows schematically what happens when such a bar, populated by stars on orbits close to the 2:1 vertical resonance, is slightly bent: the concave upper surface of the bent bar focuses the stars, concentrating them to collide with the lower surface of the bar near its end, whereas the convex lower surface of the bar de-focuses stars away from the upper surface of the bar near its ends. The net effect is thus an excess downward pressure on the ends of the bar, causing it to bend further; this positive feedback will make the bar buckle. We can thus see that the buckling phenomenon is intrinsically a collective effect which arises because stars on different orbits respond differently to perturbations in the gravitational potential. This model also once again emphasizes the importance of the 2:1 vertical resonance, and explains why bars buckle in a single bend about their mid-points.

## 3. Photometric Evidence for a Peanut/Bar Connection

As we have seen, vertical instability seems to be a generic feature of galaxy bars. What, then, is the observational evidence that thickened structures in the centres of galaxies such as peanut-shaped bulges are connected with bars in these systems? The connection between these two phenomena has proved difficult to establish. As already mentioned, vertical structure is only readily observable in edge-on galaxies, whereas the presence of a bar in a disk galaxy is only apparent if the disk is viewed fairly face-on. In seeking to overcome this problem, Bettoni & Galletta (1994) have found a barred galaxy at an intermediate inclination (NGC 4442) where the boxy structure appears to be so strong that it is still detectable even in this inclined system. As these authors point out, the photometry of NGC 4442 with it boxy isophotes and twisting major axis is remarkably similar to what is seen in Combes et al.'s (1990) simulations of peanut-shaped bars. However, photometric observations do not directly constrain the third dimension in such systems — NGC 4442 could, for example, be an entirely flat disk which just happens to have a strange isophotal structure.

Further indirect evidence for a link between peanut-shaped bulges and bars comes from photometry of galaxies like the edge-on S0 galaxy, NGC1381. This system has a very boxy bulge. It also has a strong central plateau in the luminosity distribution along its major axis, which looks very like the contribution from the flat luminosity distribution of a bar (de Carvalho & da Costa 1987). Similarly, Dettmar & Barteldrees (1990) have found thin central bar-like structures in a large number of edge-on galaxies with boxy bulges. However, none of these photometric observations provides unequivocal evidence for a bar — all of these systems could also be modelled as axisymmetric disks.

## 4. Kinematic Evidence for a Peanut/Bar Connection

Kinematic observations offer further evidence for the connection between peanut-shaped bulges and galaxy bars. For example, N-body simulations show that, unlike normal galaxy bulges, the peanut-shaped structure associated with a bar should rotate cylindrically (i.e. with azimuthal velocities that do not vary with distance from the plane of the galaxy). Observations of peanut-shaped structures in edge-on galaxies have shown that they also exhibit this property (Jarvis 1987, 1990). However, it is also quite possible to construct axisymmetric bulge models which appear peanut-shaped and rotate cylindrically (Rowley 1988).

Figure 2. Spatial distribution of stars in a set of disk galaxy models. Bar strength (parameterized by $\epsilon$) increases to the right, and viewing angle changes up the page.

Recently, it has been shown that we can use detailed kinematic observations to detect bars in edge-on galaxies directly. Useful insight into the kinematics of barred galaxies can be obtained by considering the closed orbits allowed by the potential: the collisional nature of gas forces it to follow non-intersecting closed orbits, and most stellar orbits can be interpreted as oscillations about closed orbits. The closed orbit families in barred potentials have been discussed extensively in the literature [see Sellwood & Wilkinson (1993) for a review]; the property that is important for identifying bars in edge-on galaxies is the fact that the major orbit families follow elongated paths, and that the direction of elongation changes by 90 degrees at orbital resonances. Inside any inner Lindblad resonance (ILR), for example, orbits are elongated perpendicular to the bar potential major axis; at radii between the ILR and co-rotation resonance, orbits are aligned along the bar axis, etc. These changes in the direction of elongation mean that there is a lack of non-intersecting orbits at the major resonances, and hence a deficit of material (particularly the collisional gas component) near these radii. To illustrate this point, Fig. 2 shows material on the non-intersecting orbits in a set of simulated galaxies with varying bar strengths, viewed from various directions.

When using spectra to make kinematic observations of a galaxy, the most general observable quantity is the density of material as a function of its pro-jected position and its line-of-sight velocity (as measured by the Doppler shift in its spectral lines) — the "$l - v$ diagram" in the parlance of galactic astronomers. Figure 3 shows the projected phase density of the material in the simulated galaxies of Fig. 2 which would be observable if these systems were viewed edge-on. It is apparent from this figure that the presence of bars in these galaxies ($\epsilon \neq 0$) introduces complicated structure to the observable kinematics due to the non-circular orbits and the absence of material close to the orbital resonances. This structure varies dramatically with the strength of the bar and its orientation, and so observations of such detailed kinematics in real edge-on galaxies should provide strong constraints on the properties of any bar that may lie hidden therein.

Figure 3.  Phase density of material as a function of projected radius ($x$-axis) and line-of-sight velocity ($y$-axis) for the material in the galaxy models of Fig. 2. The models were "observed" edge-on from the bottom of Fig. 2.

To search for this kinematic signature, we have obtained spectra of two edge-on galaxies with peanut-shaped bulges (NGC 5746 and NGC 5965; Kuijken & Merrifield 1995). The spectra were obtained using the ISIS two armed spectrograph on the William Herschel Telescope. The blue arm of the spectrograph was used to observe the Mg b absorption feature at 5190Å, and the red arm was used to observe either H$\alpha$ emission or the Ca absorption triplet at $\sim 8600$Å. The gaseous kinematics were derived directly from the emission line spectra, and the stellar kinematics were inferred by using the Unresolved

Figure 4. Photometry and kinematics of NGC 5746. (*a*) I-band photometry of the galaxy, with contours spaced by 0.3 magnitudes: the thick line indicates the position of the slit for the spectral observations. Line-of-sight velocity distribution, normalized to unit velocity integral, as a function of projected radius for: (*b*) the gaseous component as traced by the [NII] line at 6583Å; (*c*) the stellar component as derived from absorption lines around the Mg b feature at 5170Å; and (*d*) the stellar component as derived from absorption lines around the Ca triplet at 8600Å. The greyscale bar indicates the relative phase space density scale in both (*c*) and (*d*).

Gaussian Decomposition algorithm (Kuijken & Merrifield 1993) to deconvolve the stellar line-of-sight velocity distribution from the Doppler broadening of the absorption lines. Figure 4 shows the kinematics derived from this process for NGC 5746. It is apparent from this figure that the detailed kinematics of both the gaseous and stellar components in this edge-on galaxy show exactly the same sort of structure as the simulations in Fig. 3. Such structure cannot occur in an axisymmetric disk galaxy, and so we find unmistakeable evidence for a bar in this galaxy. Our second target with a peanut-shaped bulge, NGC 5965, revealed similar structure in both its gaseous and stellar kinematics. Further, long-slit spectra of the H$\alpha$ line in the edge-on disk galaxy NGC 2683 show the same kinematic signature (Rubin, private communication), and inspection of images of this galaxy reveal that it has a very boxy bulge.

Here, then, is the first unequivocal evidence that peanut-shaped bulges are associated with bars. Clearly, the sample size must be increased if we are to show that all peanut-shaped structures are a by-product of bar formation. It would also be interesting to look for the kinematic signature of a bar in some edge-on galaxies with non-boxy bulges. From such observations we would be able to see if any thin bars exist, or whether the buckling instability means that all bars are destined to end up as peanut-shaped structures.

**Acknowledgments.** Many thanks to Konrad Kuijken, my collaborator on this project. This research was supported by a PPARC Advanced Fellowship (B/94/AF/1840).

### Discussion

*Prof. Hawarden*: Does the excellent agreement of your absorption line kinematics at both 5000Å and 8000Å not only rule out the falsification of the observed structure by extinction but also the suggested high opacity levels in galaxies?

*Dr. Merrifield*: The galaxies that we have observed thus far are somewhat inclined to the line-of-sight ($\sim$ 5 degrees from edge-on), and the effects of even a fairly high opacity on the observed kinematics are limited in such cases (Bosma et al. 1992). However, application of the same method to galaxies which are even closer to edge-on does provide an exciting new approach to measuring the opacity of spiral galaxies. We are already seeking telescope time to make such a measurement.

*Prof. Quillen*: How difficult would it be to determine observationally whether a face-on bar is a peanut?

*Dr. Merrifield*: Once again, stellar kinematics may provide the key: the stars' line-of-sight motions perpendicular to the plane of a face-on bar offers a direct measure of the structure of the bar in the third dimension. Comparison between the stellar kinematics obtained from long slit spectra of face-on bars and the predictions of bar models should allow us to answer this question.

*Prof. Freeman*: Would you expect to see your effect in the motions of the stars in the galactic bulge?

*Dr. Merrifield*: The hotter the stellar population, the more "washed out" the structure in the line-of-sight velocity distribution. However, if we could observe sufficiently close to the plane to pick up a reasonable fraction of kinematically cooler disk stars then such observations would provide a new window on the galactic bar.

*Prof. Miller*: An unstable object cannot exist in nature. Your discussion of instability is a nice description of an object (a thin disk) that could not have formed in the first place. Instabilities in numerical models simply tell you that you've set up something that cannot exist in the real world.

*Dr. Merrifield*: I don't believe that this line of argument invalidates the use of instability analysis. The fact that all the pens on my desk are lying down rather than standing on their points can usefully be explained by the instability of the latter arrangement. Even if (as the simulations seem to show) a bar buckles as soon as it forms, the instability analysis still shows why the bar follows this evolutionary track, and provides a causal explanation for the phenomenon.